\documentclass[aps, prb, reprint, superscriptaddress, floatfix]{revtex4-2}
\usepackage{graphicx}
\usepackage{dcolumn}
\usepackage{bm}
\usepackage{physics}
\usepackage{subfigure}
\usepackage{calrsfs}
\DeclareMathAlphabet{\pazocal}{OMS}{zplm}{m}{n}
\usepackage{amsmath}
\usepackage{amsfonts}
\usepackage{amssymb}
\usepackage{mathrsfs}
\usepackage{soul}

\usepackage{xcolor}
\usepackage[colorlinks=true,linkcolor=blue,anchorcolor=red,citecolor=blue, urlcolor=blue]{hyperref}\usepackage{ulem}
\usepackage[T1]{fontenc}
\newlength{\seplinewidth}
\newlength{\seplinesep}
\colorlet{sepline}{orange}

\begin{document}
\preprint{APS/123-QED}

\title{Driven Majorana Modes: A Route to Synthetic \texorpdfstring{$p_x+ip_y$}{pdfbookmark} Superconductivity}

\author {Lingyu Yang}
\affiliation{Theoretical Division, T-4 and CNLS, Los Alamos National Laboratory, Los Alamos, New Mexico 87545, USA}
\affiliation{Department of Physics, University of Virginia, Charlottesville, Virginia, 22904, USA}

\author {Gia-Wei Chern}
\affiliation{Department of Physics, University of Virginia, Charlottesville, Virginia, 22904, USA}

\author {Shi-Zeng Lin}
\affiliation{Theoretical Division, T-4 and CNLS, Los Alamos National Laboratory, Los Alamos, New Mexico 87545, USA}
\affiliation{Center for Integrated Nanotechnologies (CINT), Los Alamos National Laboratory, Los Alamos, New Mexico 87545, USA}

\begin{abstract}
We propose a protocol to realize synthetic $p_x+ip_y$ superconductors in one-dimensional topological systems that host Majorana fermions. By periodically driving a localized Majorana mode across the system, our protocol realizes a topological pumping of Majorana fermions, analogous to the adiabatic Thouless pumping of electrical charges. Importantly, similar to the realization of a Chern insulator through Thouless pumping, we show that pumping of Majorana zero modes could lead to a $p_x + ip_y$ superconductor in the two dimensions of space and synthetic time. The Floquet theory is employed to map the driven one-dimensional system to a two-dimensional synthetic system by considering frequency as a new dimension. We demonstrate such Floquet $p_x + i p_y$ superconductors using the Kitaev $p$-wave superconductor chain, a prototypical 1D topological system, as well as its more realistic realization in the 1D Kondo lattice model as examples.  We further show the appearance of a new $\pi$ Majorana mode at the Floquet zone boundary in an intermediate drive frequency region. Our work suggests a driven magnetic spiral coupled to a superconductor as a promising platform for the realization of novel topological superconductors. 
\end{abstract}
\date{\today}
\maketitle

\section{Introduction}
\label{sec:intro}

Majorana fermions are particles that act as their own antiparticles and were originally proposed as a potential model for neural elementary particles with spin-1/2. Although it remains to be seen whether elementary particles such as neutrinos are Majorana fermions, it has been shown that emergent quasiparticles in certain quantum materials behave as Majorana fermions. In such condensed-matter systems, Majorana fermions can be viewed as collective many-body modes as the zero modes bound to vortices of a topological superconductor \cite{Kitaev_2001}, or resulting from fractionalization of elementary excitations in, e.g. quantum spin liquids \cite{balents2010spin,Savary_Balents_2017}. Importantly, the strong quantum entanglement between Majorana quasiparticles gives rise to unique properties, such as non-Abelian braiding statistics and immunity to local perturbations, making them promising candidates for building fault-tolerant quantum computers \cite{nayak2008non}. 

Numerous theoretical scenarios have been proposed for the realization of Majorana fermions in material systems~\cite{fu2008superconducting,sau2010generic, alicea2010majorana, lutchyn2010majorana,choy2011majorana,martin2012majorana,Nadi-perge2013proposal,klinovaja2013topological,vazifeh2013self}. The predominant approach is to utilize the proximity effect caused by a conventional $s$-wave superconductor to induce Majorana bound states in a heterostructure with materials endowed with a strong spin-orbit coupling~\cite{fu2008superconducting,sau2010generic, alicea2010majorana, lutchyn2010majorana,choy2011majorana}. Similar Majorana bound states can also be realized via proximity of a Kondo chain and an $s$-wave superconductor~\cite{martin2012majorana,Nadi-perge2013proposal,klinovaja2013topological,vazifeh2013self}. Despite extensive efforts in the implementation of these theoretical proposals, an unequivocal experimental demonstration of Majorana fermions remains a challenging task.

Majorana fermions also occur as localized zero-energy modes at vortex cores of a chiral triplet superconductor with $p_x + i p_y$ pairing symmetry \cite{read2000paired, Ivanov2001non-abelian, sarma2006proposal, tewari2007index}. Quantum computation can be performed by braiding Majorana fermions through  controlled motions of vortices \cite{lian2018topological}. However, the $p_x + i p_y$ superconducting pairing, which spontaneously breaks the time-reversal symmetry, is rather rare. To date,  only a few superconductors, such as $\rm{Sr_{2}RuO_{4}}$ \cite{rice1995sr2ruo4, mackenzie2003superconductivity, raghu2010hidden, wang2013theory}, $\rm{Au_{2}Pb}$ \cite{xing2016superconductivity}, and doped $\rm{Be_{2}Se_{3}}$ systems \cite{wang2012coexistence, li2017origin, zhang2011superconducting, zareapour2012proximity, wang2013fully} etc., have been suggested to exhibit this pairing symmetry. Strontium ruthenate $\rm{Sr_{2}RuO_{4}}$, which has for a long time been considered a promising candidate for the chiral superconductor, has been ruled out by recent nuclear magnetic resonance measurement \cite{Pustogow_Luo2019}, thus further shrinking the candidate pool. Therefore, it remains a challenge to realize the $p_x + i p_y$ pairing symmetry in superconductors and it is highly desirable to find an alternative approach.

The intimate relationship between chiral $p$-wave superconductors and Majorana bound states suggests an alternative route to engineering $p_x + i p_y$ superconductors based on existing topological systems that support Majorana fermions. In particular, a dynamical approach similar to Thouless pumping could lead to the emergence of synthetic $p_x + i p_y$ pairing in a periodically driven 1D system. Theoretically, such synthetic quantum systems can be described using the Floquet formalism, which introduces an effective stationary Hamiltonian, dubbed the ``Floquet Hamiltonian'', $H_F =  i\hbar \log U(T) / T$, where $U(T)$ is the time evolution operator of the periodically driven system in a full period $T$. Indeed, in the last decade, Floquet engineering has emerged as a promising route to inducing new phases or modifying existing phases of quantum materials \cite{PhysRevLett.121.107201,PhysRevB.100.220403,Mentink2015,PhysRevB.96.014406,Mikami2016,wang_observation_2013,mciver_light-induced_2020,oka2019floquet,RevModPhys.93.041002,Shan_Ye_Chu_Lee_Park_Balents_Hsieh_2021,PhysRevB.103.064508, Kumar_Banerjee_Lin_2022}. For example, the Floquet method has been used to realize the Chern insulator in graphene \cite{PhysRevB.79.081406, Kitagawa2011transport, delplace2013merging, grushin2014floquet, gomezleon2014engineering, PhysRevB.103.064508, Kumar_Banerjee_Lin_2022, PhysRevB.105.L180414, plekhanov2017floquet,Park_Lee_Jang_Choi_Park_Jung_Watanabe_Taniguchi_Cho_Lee_2022}. 


In this work, we apply the Floquet method to engineer a synthetic $p_x + i p_y$ superconductivity. Our idea is based on the pumping of Majorana fermions under periodic drive of a topological chain, analogous to the Thouless pumping mechanism for the emergence of a synthetic Chern insulator. We first demonstrate the proposed scenario using the Kitaev $p$-wave superconductor model under a time-dependent chemical potential. An effective 1+1D Floquet Hamiltonian, where the extra $y$ direction represents the Floquet frequency, exhibits a $p_x + i p_y$ pairing symmetry. Next, we investigate the driven Majorana mode mechanism on a more realistic physical model built from a Kondo chain in proximity to an $s$-wave superconductor. The nontrivial topological Floquet $p_x + i p_y$ superconductor manifests itself further in the emergence of a new Majorana $\pi$ mode at the zone boundary of the Floquet-Brillouin zone for intermediate driving frequencies.

The remainder of the paper is organized as follows. In Sec.~II, we introduce the idea of Majorana pumping leading to the emergence of 1+1D synthetic $p_x + i p_y$ pairing, as well as its implementation in the Kitaev $p$-wave superconducting chain model. The relevant Floquet theory is also briefly reviewed. The realization of Majorana bound states in the 1D Kondo lattice model is then discussed in Sec.~III. We show that the pumping of Majorana fermions can be achieved by a rotating magnetic spiral. A Floquet theory is developed to describe the periodically driven Kondo chain coupled to the $s$-wave superconductor.  In Sec. IV, we study the Majorana $\pi$ modes for both the driven Kitaev and Kondo lattice models. The paper ends with a conclusion and outlook in Sec. V.

\section{Pumping majorana fermions}


We first review the mechanism of Thouless pumping, which is the construction of a 1+1D topological phase from periodic pumping of charges in the 1D chain~\cite{Asboth2016}. We then propose a similar mechanism for the emergence of a synthetic $p_x + i p_y$ superconductor via the pumping of Majorana zero modes (MZMs) in 1D topological chains.
Consider the 1D Rice-Mele Hamiltonian with a periodically modulation of the hopping and on-site potential: 
\begin{eqnarray}
	& & \mathcal{H}_{\rm RM} = \sum_j \left\{ [1 + \cos(\omega t)] c_{A,j}^\dagger c^{\,}_{B, j} + c^\dagger_{B, j} c^{\,}_{A, j+1} + \mbox{h.c.} \right\} \nonumber \\
	& & \qquad + \sin(\omega t) \sum_j \left( c^{\dagger}_{A, j} c^{\,}_{A, j} - c^{\dagger}_{B, j} c^{\,}_{B, j} \right).
\end{eqnarray}
Here, we assume the adiabatic limit with $\omega \ll 1$. The effect of the periodic driving can be understood as follows. In the first half of each period, say $\omega t \in [0, \pi]$, sublattice B has a lower on-site potential compared with that of sublattice A. As a result, electrons tend to accumulate in the B sublattice. In the second half of each period, hopping on half of the nearest-neighbor bonds is suppressed. As the on-site potential also changes sign between the two sublattices, electrons are encouraged to move from B- to the A-sublattice. The net effect is a coherent one-way transport of electrons along the chain, a mechanism similar to a classical Archimede's screw. In fact, it can be proved that exact one electron is pumped from one end of the chain to the other in one driving period. To see how the above periodic drive leads to the formation of an effective Chern insulator, we perform the spatial Fourier transform to the Rice-Mele Hamiltonian: $\mathcal{H}_{\rm RM} = \sum_{k_x} \bm c^{\dagger}_{k_x} \cdot  H_{\rm RM}(k_x; \omega t) \cdot \bm c^{\,}_{k_x}$, where $\bm c^{\,}_{k_x} = (c_{A, k_x}, c_{B, k_x})^T$, and the one-particle Hamiltonian reads:
\begin{equation}
H_{\rm{RM}}
=\left(1+\cos(\omega t)+\cos k\right)\hat{\sigma}_x + \sin k\, \hat{\sigma}_y + \sin(\omega t)\hat{\sigma}_z,
\label{RM_k}
\end{equation}
where the Pauli matrices $\hat{\sigma}$ are defined in the sublattice basis. In the slow driving limit, the electrons are assumed to stay adiabatically in the ground state of the instantaneous Hamiltonian. Moreover, since the Hamiltonian and the corresponding ground state only depend on the $\omega t$ module $2\pi$ in the adiabatic limit, $\omega t$ can be treated as an effective wave vector, to be labeled by $k_y$, defined in a Brillouin zone $[0, 2\pi)$. After performing the rotation from $(\hat{\sigma}_x, \hat{\sigma}_y, \hat{\sigma}_z)$ to $(\hat{\sigma}_z, \hat{\sigma}_x, \hat{\sigma}_y)$, the adiabatic Rice-Mele Hamiltonian can be recast to a standard form considered by Qi, Wu, and Zhang~\cite{PhysRevB.74.085308}
\begin{equation}
H_{\rm{QWZ}}=\sin k_x \hat{\sigma}_x + \sin k_y \hat{\sigma}_y +\left(1 + \cos k_x + \cos k_y\right)\hat{\sigma}_z.
\label{QWZ}
\end{equation}
Importantly, this Hamiltonian is shown to exhibit a quantum anomalous Hall effect, and is characterized by a topological Chern number $C = \pm 1$.  For a finite system with periodic boundary condition along the $y$ -axis while open boundary condition along the $x$-axis, the Qi-Wu-Zhang model is reduced to Rice-Mele model labeled with different $k_y$. Thus, there exist edge modes localized at the left and right ends of the $x$ axis for an open chain of the Rice-Mele model, and their energy varies with $k_y$.

\begin{figure}[t]
\includegraphics[width=85mm]{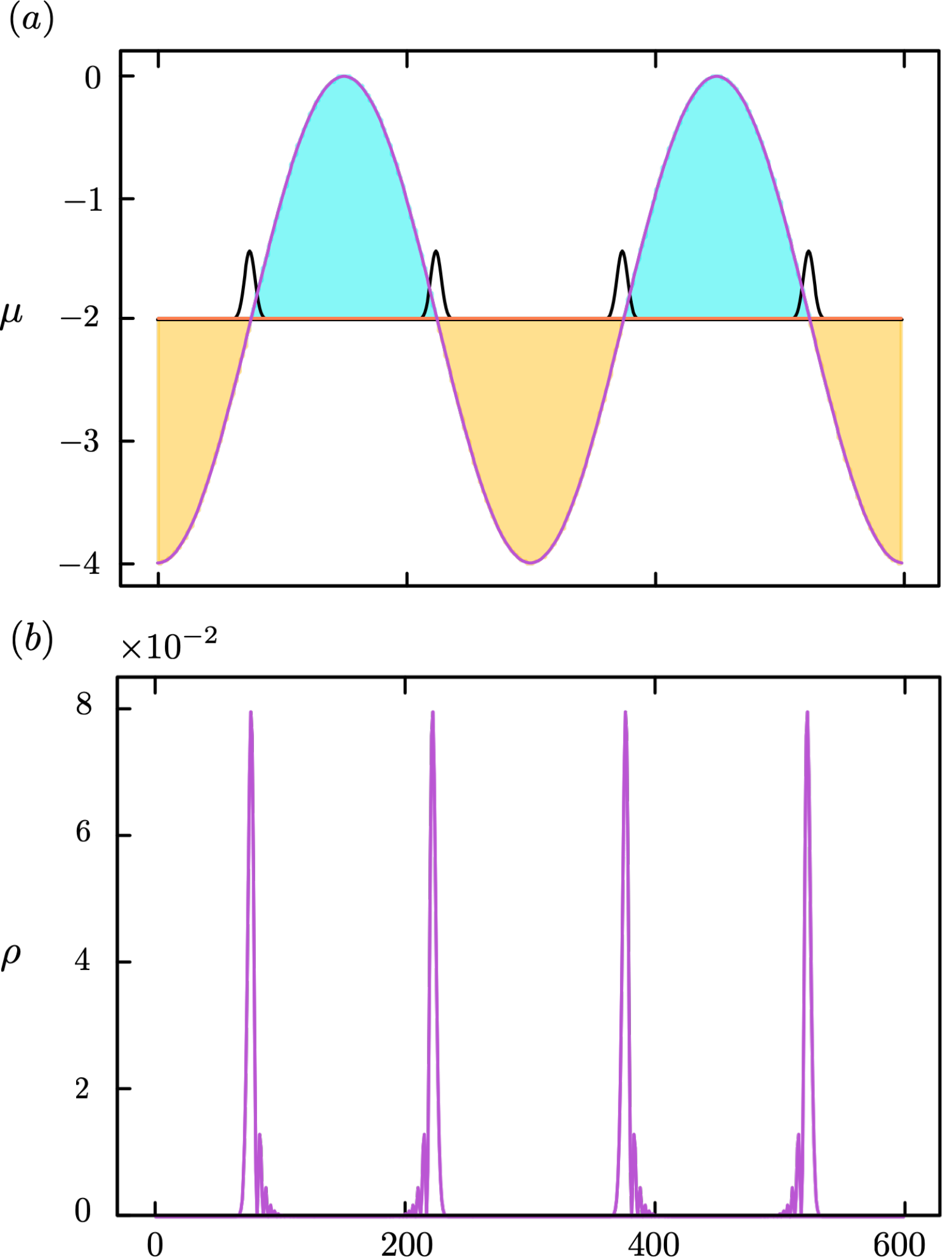}
\caption{\label{mod_mu_wf}(a) Spatial profile of a modulated chemical potential $\mu_j = \mu_{0}+ \delta\mu \cos(qj+\varphi)$, with $\mu_0 = -2t$, $\delta\mu = 2t$, and $q = \pi / 150$ for the modified Kitaev chain. Topological superconducting domains (blue regions) are stabilized in regions where $-2t<\mu_j<2t$, while normal superconducting domains (gold regions) correspond to regions with $\mu_j < -2 t$. MZMs appear at the interface of the topological and normal superconducting domains, as labelled by the black curves. (b)~Numerical calculations of the spatial profile of each MZMs whose peaks are localized at the interfaces as predicted.}
\end{figure}

\subsection{Modulated Kitaev model}

The above mechanism of Thouless pumping suggests a dynamical approach to engineer or design synthetic topological phases. In particular, here we show that similar periodic pumping of MZMs could lead to the emergence of a synthetic $p_x + i p_y$ superconductor. To this end, we first consider a modified Kitaev $p$-wave superconductor model with a site-dependent chemical potential.
\begin{equation}
\mathcal{H}_{\rm{K}}=\sum_{j}\left( -t\,c^{\dag}_{j}c^{\,}_{j+1} -\mu_j \, c^{\dag}_{j}c^{\,}_{j} + \Delta\,c^{\dag}_{j+1}c^{\dag}_{j} + \mbox{h.c.}\right),
\label{kitaev}
\end{equation}
where $t$ is the nearest-neighbor hopping coefficient, $\mu_j$ is the on-site chemical potential, and $\Delta$ is the superconducting order parameter, which is assumed to be real. The Kitaev chain with a uniform chemical potential $\mu$ is topological when it lies in the interval $-2t<\mu<2t$. The bulk-boundary correspondence of this topological phase is manifested in the occurrence of MZMs localized at the ends of an open chain. In the presence of spatially modulated chemical potential, both topological and non-topological superconducting domains coexist in the system. An interface of the two types of domain occurs at the point where $\mu_j$ crosses $\pm 2t$. Importantly, a MZM is bound to such an interface separating topological superconducting domains from the normal ones.

Specifically, we consider a sinusoidal modulation of the chemical potential $\mu_j = \mu_{0}+ \delta\mu \cos(qj+\varphi)$, where $\mu_0$ is the average chemical potential, $\delta\mu$ is the modulation amplitude, and the wave vector $q$ determines the modulation period. Fig.~\ref{mod_mu_wf}(a) shows the spatial profile of a chemical potential with $\mu_0 = -2t$, $\delta\mu = 2t$, and $q = \pi / 150$. The two types of superconducting domain, topological versus normal, are shown in different colors. MZMs are obtained at the points where $\mu_j$ crosses $-2t$. Fig.~\ref{mod_mu_wf}(b) shows the probability densities of the MZMs, indicating the highly localized nature of such zero mode states. 

Importantly, pumping of such localized MZMs can be achieved by slowly moving the interfaces or domain-walls across the system, for example, by introducing a traveling sinusoidal waveform for the chemical potential 
\begin{eqnarray}
	\label{mu_j}
	\mu_j(t) = \mu_{0}+ \delta\mu \cos(qj+\omega t).
\end{eqnarray}
In the adiabatic limit, the speed of the traveling MZMs is given by $v = \omega/q$. In the following, we demonstrate the emergence of a $p_x + i p_y$ superconductor from the pumping of MZMs using the Floquet method.

\subsection{Floquet theory}
We first consider a general single-particle time-periodic Hamiltonian $H(t)=H(t+T)$. Based on the analogy with a spatially periodic potential $V(x) = V(x+a)$, we are interested in the so-called Floquet state $\ket{\phi_{\alpha}}=e^{-i\epsilon_{\alpha}t}\ket{\theta_{\alpha}(t)}$, where $\epsilon_{\alpha}$ is the quasienergy, and $\ket{\theta_{\alpha}(t)} = \ket{\theta_{\alpha}(t + T)}$ is a state with periodicity $T$. The time-dependent Schr\"odinger equation becomes an effective Schr\"odinger eigenvalue equation for the Floquet states
\begin{eqnarray*}
H_{\rm F}\ket{\theta_{\alpha}(t)}=\epsilon_{\alpha}\ket{\theta_{\alpha}(t)},
\end{eqnarray*}
where $H_{\rm F}=H(t)-i\partial_t$ is called the Floquet Hamiltonian. Expanding the Floquet state in a Fourier series, $\ket{\theta_{\alpha}(t)}=\sum_m e^{im\omega t}\ket{\Phi^m_{\alpha}}$ and $H(t)=\sum_m e^{-im\omega t}H_m$, the effective Schr\"odinger equation becomes \cite{oka2019floquet}
\begin{eqnarray*}
\begin{centering}
   \sum_{m}H^{n,m}_{\rm F}\ket{\Phi^m_{\alpha}} = \epsilon_{\alpha}\ket{\Phi^m_{\alpha}}
\end{centering}
\end{eqnarray*}
where $ H^{n,m}_{\rm F} = \frac{1}{T}\int^{T}_{0} H(t)e^{-i(n-m)\omega t}dt-n\omega\delta_{mn}$ denotes the Fourier expansion of the Floquet Hamiltonian. \par

\begin{figure*}[t]
\includegraphics[width=188mm]{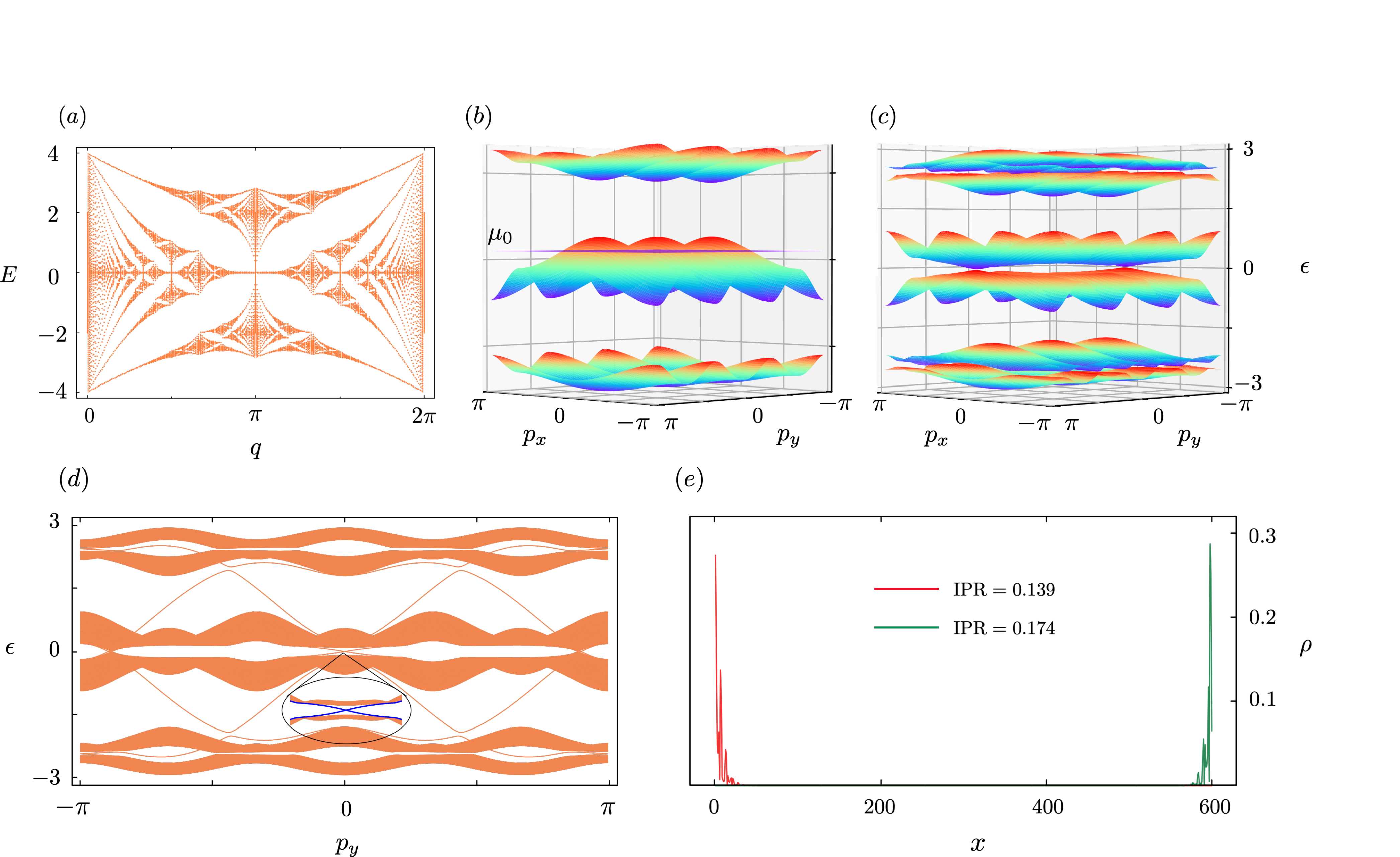}
\caption{\label{mod_kitaev}(a)~Quasienergy spectrum of the normal state of modulated Kitaev model exhibiting the Hofstadter butterfly. (b)~Quasienergy band structures of the three-sites model in the normal state. (c)~Quasienergy band structures of the three-sites model in the superconducting state. The Chern numbers for each band (from bottom to top) are $(-1, +1, +1, -1, -1, +1)$. (d)~Quasienergy spectrum of $H_{F}$ using the periodic boundary conditions along the frequency direction and the open boundary condition along the $x$ direction. (e)~Spatial profile and IPRs of the chiral MZMs at $p_{y}=0.2$. Parameters: $t=1.0$, $\mu_0=0.2$, $\delta\mu=2.0$, $\Delta=0.1$.}
\end{figure*}

We next apply the Floquet theory to the modulated Kitaev model~(\ref{kitaev}) with a space-time dependent chemical potential in Eq.~(\ref{mu_j}). It is more convenient to first work out the Fourier transformation in second-quantized form
\begin{equation}
\begin{aligned}
\mathcal{H}_{\rm{K}}(t)=\sum_{j}&\left[-t\,c^{\dag}_jc_{j+1}+|\Delta|\,c^{\dag}_{j+1}c^{\dag}_{j}+\mbox{h.c.}\right.\\ &\left.-\mu_0\,c^{\dag}_{j}c_{j}-\delta\mu\cos(qj+\omega t) c^{\dag}_{j}c_{j}\right]
\end{aligned}
\label{modkitaev}
\end{equation}
The Floquet Hamiltonian diagonal in the frequency domain comes from the time-independent terms and the Fourier transform of the $-i \partial_t$ operator
\begin{equation}
\begin{aligned}
   \mathcal{H}^{mm}_{\rm F}=\sum_{j}&\left(-t\,c^{\dag}_{j,m}c_{j+1,m}+|\Delta|\,c^{\dag}_{j+1,m}c^{\dag}_{j,m}+h.c.\right.\\
   &\left.-\mu_{0}\,c^{\dag}_{j,m}c_{j,m} -m\omega\, c^{\dag}_{j,m}c_{j,m}\right)
\end{aligned}
\label{modkitaev1}
\end{equation}
The traveling sinusoidal wave for the chemical potential contributes to the Floquet Hamiltonian which is off-diagonal in frequency domain
\begin{equation}
	\mathcal{H}^{m,m\pm1}_{\rm F}=-\sum_{j} \frac{\delta\mu}{2} \left(e^{-iqj}c^{\dag}_{j,m}c_{j,m+1} + e^{iqj}c^{\dag}_{j,m}c_{j,m-1}\right)
\label{modkitaev2}
\end{equation}
These off-diagonal terms describe transitions between different frequency components, thus does not conserve energy. All other entries of $\mathcal{H}^{n,m}_{\rm F}$ except those in Eqs. \eqref{modkitaev1} and \eqref{modkitaev2} are zero.



By expressing the second quantized Floquet Hamiltonians in the form $\mathcal{H}^{m, n}_{\rm F} = \sum_{ij} \bm c^{\dagger}_{i,m} H^{\,}_{im, jn} \bm c^{\,}_{j,n}$, where $\bm c^{\,}_{j,n} = (c^{\,}_{j,n}, c^{\dagger}_{j, n})^T$, 
the resultant single-particle Hamiltonian $H_{im, jn}$ thus describes an effective 2D tight-binding model of superconductivity in which the Fourier component index $m$ becomes a synthetic $y$-axis in the 2D lattice.  The hopping strength along the frequency axis carries a phase $\exp(-i q j)$ that depends on the real space coordinates. Therefore, each plaquette carries an effective magnetic flux of $q$. The effective model Eqs. \eqref{modkitaev1} and \eqref{modkitaev2} is the same as the Hofstadter model on a square lattice with a superconducting pairing term, and $\mu_0+m\omega$ is the on-site chemical potential term. Assuming that $\omega$ is small, we may omit the last term in Eq. \eqref{modkitaev1} such that the translational invariance of the Hamiltonian associated with the magnetic translation group is restored.\par

We first consider the normal state of the effective 2D Hamiltonian by turning off the superconducting pairing, i.e. $\Delta=0$, in this 2D model. The quasienergy as a function of $q$, shown in Fig.~\ref{mod_kitaev}(a), exhibits the characteristic Hofstadter butterfly. The normal state has topological bands at different $q$. We choose $q=2\pi/3$ as an example to study the emergence of $p_x+ip_y$ superconductivity when a pairing term is present. In this commensurate modulation with three lattice sites in an extended unit cell, the Fourier-transformed Hamiltonian in the basis of $\Psi=(c_{1,p_x,p_y}, c_{2,p_y}, c_{3,p_x, p_y})^{T}$ is
\begin{widetext}
\begin{eqnarray}
\label{3x3H}
   H=\left(                 
   \setlength{\arraycolsep}{1.2pt}
 \begin{array}{ccc}   
   -\mu_0-\delta\mu\cos\left(q+p_y\right) & -t & -t\,e^{ip_x}\\[2.5pt]  
   -t & -\mu_0-\delta\mu\cos\left(2q+p_y\right) & -t       \\[2.5pt]   
   -t\,e^{-ip_x} & -t & -\mu_0-\delta\mu\cos\left(3q+p_y\right) \\
 \end{array}
 \right)
\end{eqnarray}
\end{widetext}
The band structures obtained from diagonalization of the $3\times 3$ matrix is shown in Fig.~\ref{mod_kitaev}(b). We remark that here we choose $q = 2\pi/3$ to make analytical calculation possible. The fact that there are only three sites in each topological and normal segments seems beyond the intuitive picture of well-defined domain walls with localized MZMs. Nonetheless, despite finite overlap of Majorana wave functions, as demonstrated below, the mechanism of Majorana pumping remains valid even for a small unit cell with a large $q$. \par

Next we turn on the superconducting pairing. The $p_x$-pairing Hamiltonian in momentum space is given by
\begin{equation}
   \begin{aligned}
         \mathcal{H}_{\Delta} =\Delta &\left( \,c^{\dag}_{1,p_x,p_y}c^{\dag}_{3,-p_x,-p_y}e^{ip_x} + c^{\dag}_{2,p_y}c^{\dag}_{1,-p_x,-p_y}\right.\\
         &\left.+\,c^{\dag}_{3,p_x,p_y}c^{\dag}_{2,-p_y} + \mbox{h.c.} \right)
   \end{aligned}
\end{equation}
The presence of pairing doubles the number of bands due to the particle-hole symmetry. The energy bands obtained from the diagonalization of the extended $6\times 6$ matrix is shown in Fig. \ref{mod_kitaev}(c). As posited above, the pumping of Majorana fermions gives rise to topological superconducting states with chiral $p$-wave pairing. To verify this, we first compute the Chern number of these quasiparticle bands using the formula~\cite{JPSJ.74.1674}
\begin{equation}
   C_{n}=\frac{1}{2\pi i}\int_{\textit{T}^2} d^2k F_{12}(k)
\label{chern}
\end{equation}
where $F_{12}(k)\equiv\partial_{1}A_{2}(k)-\partial_{2}A_{1}(k)$ is the Berry curvature, $A_{\mu}\equiv \langle n(k) | \partial_{\mu} |n(k) \rangle $ is the Berry connection, and $\ket{n(k)}$ is the normalized wave-functions of the $n$th band \cite{PhysRevLett.49.405, berry1984quantal, PhysRevLett.51.2167}. The Chern numbers of the six bands in Fig.~\ref{mod_kitaev}(c) are, from bottom to top, $(-1, +1, +1, -1, -1, +1)$.

\par


The topological nature of these quasiparticle bands also manifests itself in the emergence of a pair of chiral MZMs localized at open ends through the bulk-boundary correspondence. The existence of such chiral MZM bands crossing the zero energy is shown in Fig.~\ref{mod_kitaev}(d) by applying the open boundary condition along the spatial $x$ direction. We further plot the probability densities of the in-gap states, $\rho_i\equiv |\psi_n(i)|^2$, and compute their inverse participation ratios (IPRs) defined as $\sum_{i}|\psi_n(i)|^4/(\sum_{i}|\psi_n(i)|^2)^2$ to check the localization of the MZMs, see Fig.~\ref{mod_kitaev}(e).

\subsection{The emergence of \texorpdfstring{$p_y$}{pdfbookmark} pairing}
\begin{figure}[t]
\includegraphics[width=85mm]{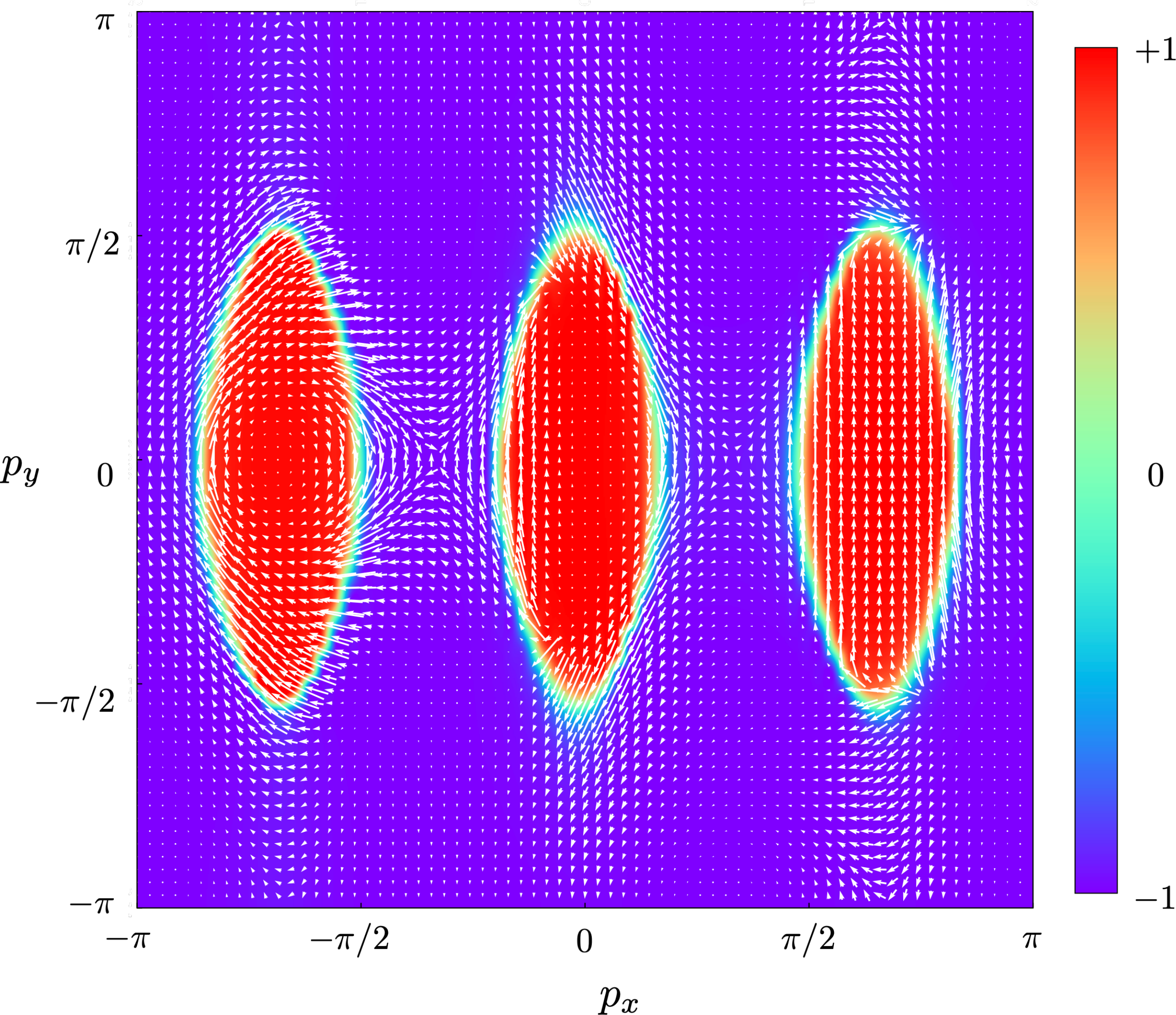}
\caption{\label{vortex} Vortex structures of $\hat{n}$ at $E_{2}=0$, which indicates that the pairings of the modulated Kitaev model are of the form of $p_x+ip_y$.}
\end{figure}

Next we demonstrate that the quasiparticle bands not only are topological, but also exhibit a chiral $p_x + i p_y$ superconductivity. To this end, we first express the $p_x$-pairing Hamiltonian of the original Kitaev chain in the Bogoliubov quasiparticle basis $(d_{1,p_x,p_y}, d_{2,p_x,p_y}, d_{3,p_x, p_y})^{T}$. In particular, after projecting to the second band of the normal-state Hamiltonian~(\ref{3x3H}) where the Fermi surface is located, we obtain
\begin{equation}
    \mathcal{H}_{\Delta} \rightarrow \Delta f(p_x,p_y) \, d^{\dag}_{2,p_x,p_y}d^{\dag}_{2,-p_x,-p_y} + \mbox{h.c.}
\end{equation}
where the form factor, $f(p_x, p_y)$, originated from the projection to the $d_{2}$ band, depends both on $p_x$ and $p_y$. Now the effective Hamiltonian is a $2\times 2$ matrix in the particle-hole basis $\Psi_{P}=(d_{2,p_y}, d^{\dag}_{2,-p_y})^{T}$,
\begin{eqnarray*}
  & &  H_{\rm eff} =
   \left(                
   \begin{array}{cc}   
       E_{2} & \Delta f(p_x,p_y)\\  
       \Delta f^{*}(p_x,p_y) & -E_{2}\\  
   \end{array} 
\right) \\
& & \quad \,\,\,\, = \mathcal{E}(p_x,p_y) \, \hat{n}(p_x, p_y) \cdot \vec{\sigma},
\end{eqnarray*}
where $E_2$ is the eigen-energy of the second band. As shown in the second equality above, in terms of the Pauli matrices representation, this Hamiltonian is characterized by a unit vector $\hat{n}(p_x, p_y)$ defined in the Floquet-Brillouin zone.  A vortex structure in $\hat{n}$ exists at $E_{2}=0$ when $\Delta f(p_{x}, p_{y})$ has a form of $p_x+ip_y$ pairing symmetry. This is indeed the case, as we plot the unit vector $\hat{n}$ in Fig.~\ref{vortex}. In fact, the Chern number is just the skyrmion number associated with the $\hat{n}$ texture in the Brillouin zone, which we verified explicitly.

\section{Driven Kondo lattice chain}
The Kitaev $p$-wave superconductor is a minimum model demonstrating the existence of MZMs, and our generalization of a traveling sinusoidal-modulated chemical potential provides a similarly minimum model of how pumping the MZMs leads to the emergence of a synthetic $p_x + i p_y$ superconductivity. As discussed in Sec.~\ref{sec:intro}, several realizations of the Kitaev model in condensed matter systems have been proposed in the past \cite{sau2012realizing, Leijnse2012parity, dvir2023realization}. In this section, we focus on the scenario of realizing localized MZMs in a 1D Kondo lattice model in proximity to an $s$-wave superconductor~\cite{martin2012majorana,Nadi-perge2013proposal,klinovaja2013topological,vazifeh2013self} as a physical realization of our proposed scenario discussed in the previous sections. 

\begin{figure*}[t]
\includegraphics[width=175mm]{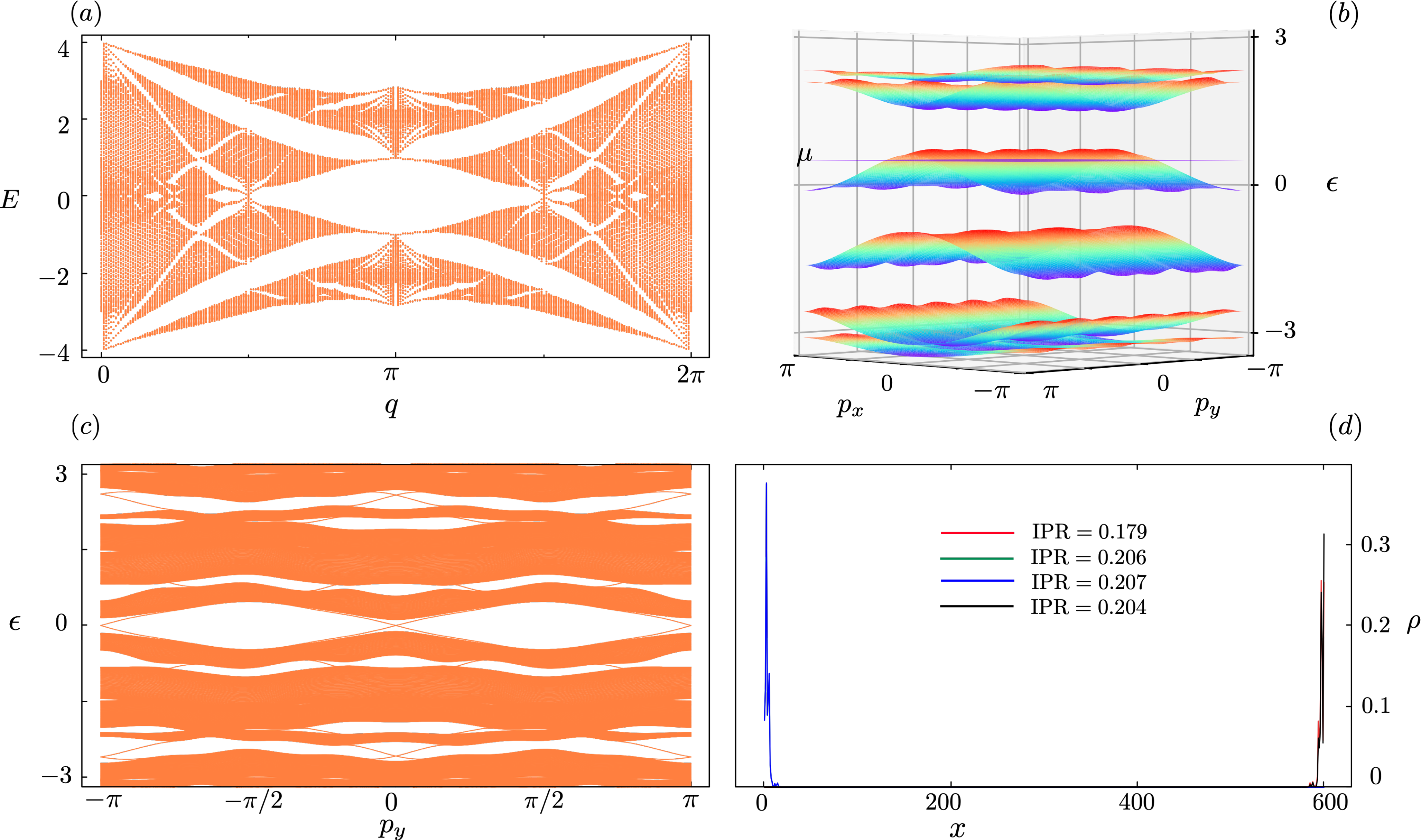}
\caption{\label{spin}(a) Hofstadter butterfly of the spin model in the normal state; (b) The quasienergy band structures of the three-site model in the normal state; (c) quasienergy spectrum of $H_{\mathrm{KL}}$ in the superconducting state using periodic boundary conditions along the time axis and open boundary conditions along the $x$ axis. (d) Spatial profile and IPRs of the chiral MZMs at $p_{y}=\pm3.0$ and $p_{y}=\pm0.08$. Parameters: $t=1.0, \mu=\Delta=b=0.5, J=2.0, B=0.1$.}
\end{figure*}
 
The Kondo lattice model coupled to the $s$-wave superconductivity is described by the Hamiltonian:
\begin{equation}
\begin{aligned}
   \mathcal{H}_{\rm{KL}}&=\sum_{j,\sigma}\left[-t\,c^{\dag}_{j\sigma}c_{j+1\sigma}+\Delta\,c^{\dag}_{j\uparrow}c^{\dag}_{j\downarrow}
           -\mu\, c^{\dag}_{j,\sigma}c_{j,\sigma}+h.c.\right]\\
           &-\sum_{j,\alpha,\beta}Jc^{\dag}_{j\alpha}(\vec{S_{j}} \cdot \vec{\sigma})_{\alpha\beta}c_{j\beta}
\end{aligned}
\label{KLM}
\end{equation}
where the last term describes the Kondo coupling between the magnetic spiral and conduction electrons, $j$ labels sites along the chain, and $\alpha, \beta$ are spin indices. Here we assume a spiral configuration for the local spins:
\[
	\vec{S_{j}}=(0,\ -\sin(qj+\varphi),\ b\cos(qj+\varphi)),
\]
where $b \in [0, 1)$ is a parameter determining the ellipticity. A traveling spiral, similar to Archimedes' screw, is then realized by a phase $\varphi = \omega t$, which increases linearly with time. To investigate the low-energy physics of this model, we first make a gauge transformation by aligning the quantization axis of electron spin with the local moment. More details can be found in Appendix~\ref{app:rotation}. The Kondo coupling term then becomes
\begin{equation*}
\setlength\abovedisplayskip{6pt}
   \begin{aligned}
       J\,U^{\dag}_{j}(\vec{S_{j}} \cdot \vec{\sigma})U_{j}&=J\sqrt{b^2\cos^{2}(qj+\varphi)+\sin^{2}(qj+\varphi)}\,\sigma_{z}\\
       &\equiv \widetilde{J}(b,j,\varphi)\sigma_{z}.
   \end{aligned}
   \end{equation*}
\hspace*{\fill}\\
In the case when $\widetilde{J}(b,j,\varphi)\approx|\mu|\gg t,\ \Delta$, and $\mu>0$, only the down-spin bands near the Fermi level are important. We next perform the Schrieffer-Wolff transformation to project out the up-spin bands, $e^{-iS}H_{\rm{KL}}e^{iS}$, and obtain the low-energy Hamiltonian $ \widetilde{H}_{\rm{eff}}=H_{\rm{KL}}+[H_{\rm{KL}}, iS]$. By keeping the terms linear in $t$ and $\Delta$, we find
\begin{widetext}
\begin{equation}
    \mathcal{H}^{\rm{eff}}_{\rm{KL}}
       =\sum_{j}\Bigg[\left(-\mu+\widetilde{J}\left(b,j,\varphi\right)
\right)\overline{c}^{\dag}_{j\downarrow}\overline{c}_{j\downarrow} -t\,\alpha_{j}\,\overline{c}^{\dag}_{j\downarrow}\overline{c}_{j+1\downarrow}
-\left( \frac{1}{2J(b,j,\varphi)} + \frac{1}{2\mu}
\right)
t\,\Delta\,\beta_{j}\, \overline{c}_{j\downarrow}\overline{c}_{j+1\downarrow} + \mbox{h.c.}
\Bigg]
\end{equation}
\end{widetext}
For the details on the Schrieffer-Wolff transformation and the definition of $\alpha_j$, $\beta_j$, see Appendix~\ref{app:wstransform}. This low energy effective Hamiltonian is a modified Kitaev $p$-wave model. Since the elliptical spiral term becomes part of the chemical potential in the effective model, we may drive the magnetic spiral to change the chemical potential and to drive the system. For a circular spiral, $b=1$, $\widetilde{J}\left(b,j,\varphi\right)$ becomes site independent, which justifies the introduction of the elliptical magnetic spiral.  \par

For a periodic spiral, the Hamiltonian is periodic in both the $x$ axis and $\varphi$, and the Hamiltonian lives in a compact 2D manifold. The Chern number can be calculated using Eq. \eqref{chern}. It is pointed out in Ref. \cite{PhysRevB.98.235116} that if one performs a rotation of electron spins around the $x$-axis by $\pi$, it yields $\mathfrak{R}_{x}(\pi)H(q,\varphi)\mathfrak{R}^{-1}_{x}(\pi) = H(q, \varphi + \pi)$, which indicates that the edge modes occur in pairs at $\varphi$ and $\varphi+\pi$. Due to the bulk-edge correspondence, the Chern number in this system must be even. However, if one performs a global rotation of the electron spins along the $z$-axis by $\pi$, one finds $\mathfrak{R}_{z}(\pi)H(q,\varphi)\mathfrak{R}^{-1}_{z}(\pi) = H(-q, -\varphi)$, which indicates that the Chern number is odd under $q\rightarrow -q$. Thus, by breaking the $\pi$-spin rotational symmetry along the $x$ axis, one may enrich the system from the $2\mathbb{Z}$ to the $\mathbb{Z}$ topological class. This can be achieved by introducing a Zeeman field, $-B\sum_{j,\alpha, \beta}c_{j\alpha}^{\dag}\hat{\sigma}^{z}_{\alpha \beta}c_{j\beta}$, along the $z$-axis. \par

Uppon setting $\varphi \rightarrow \omega t$ and performing the Floquet transformation, we derive an effective 2D Hamiltonian for the driven Kondo lattice model. The nonzero entries of $\mathcal{H}^{nm}_{\rm{KL}}$ are
\begin{widetext}
\begin{equation}
    \mathcal{H}^{mm}_{\rm{KL}}=\sum_{j\sigma}\left(-t\, c^{\dag}_{j,m,\sigma}c_{j+1,m,\sigma}+\Delta\, c^{\dag}_{jm\uparrow}c^{\dag}_{jm\downarrow} + \mbox{h.c.}
   -\mu\,c^{\dag}_{jm\sigma}c_{jm\sigma} -m\omega\,c^{\dag}_{j,m}c_{j,m}\right)
   -B\sum_{j,\alpha,\beta}c_{jm\alpha}^{\dag}\hat{\sigma}^{z}_{\alpha \beta}c_{jm\beta}
\label{KLM1}
\end{equation}

\begin{equation}
        \mathcal{H}^{m,m\pm1}_{\rm{KL}}=
        -\frac{J}{2}\sum_{j} \Bigg[ b\,e^{\mp iqj}c^{\dag}_{j,m\uparrow}c_{j,m\pm 1,\uparrow} \mp e^{\mp iqj}c^{\dag}_{j,m\uparrow}c_{j,m\pm 1,\downarrow} 
        \pm e^{\mp iqj}c^{\dag}_{j,m\downarrow}c_{j,m\pm 1\uparrow} - b\,e^{\mp iqj}c^{\dag}_{j,m\downarrow}c_{j,m\pm 1,\downarrow} \Bigg]
\end{equation}
\end{widetext}
The $-m\omega\, c^{\dag}_{j,m}c_{j,m}$ term in Eq.~\eqref{KLM1} can again be omitted using the same arguments in Eq.~\eqref{modkitaev1}. We apply the periodic boundary condition for both the spatial and the time directions, with the momentum labeled as $p_x$ and $p_y$ respectively. Again, we first turn off the superconducting pairing and plot the quasienergy spectrum at different $q$, as shown in Fig.~\ref{spin}(a). The quasienergy band structures of the corresponding three-sites model at $q=2\pi/3$ are shown in Fig.~\ref{spin}(b). The value of $\mu$ is chosen such that it sits in the third highest band to ensure that the parent compound is in a metallic state. The quasienergy bands of the effective 2D model is shown in Fig.~\ref{spin}(c). We found two pairs of chiral MZMs near the zero energy level. The total Chern number at this level is calculated to be two, which is expected from the bulk-edge correspondence. The spatial profiles and IPRs of the two pairs of chiral MZMs are calculated to check their localization, as shown in Fig.~\ref{spin}(d).

\begin{figure}[t]
\includegraphics[width=85mm]{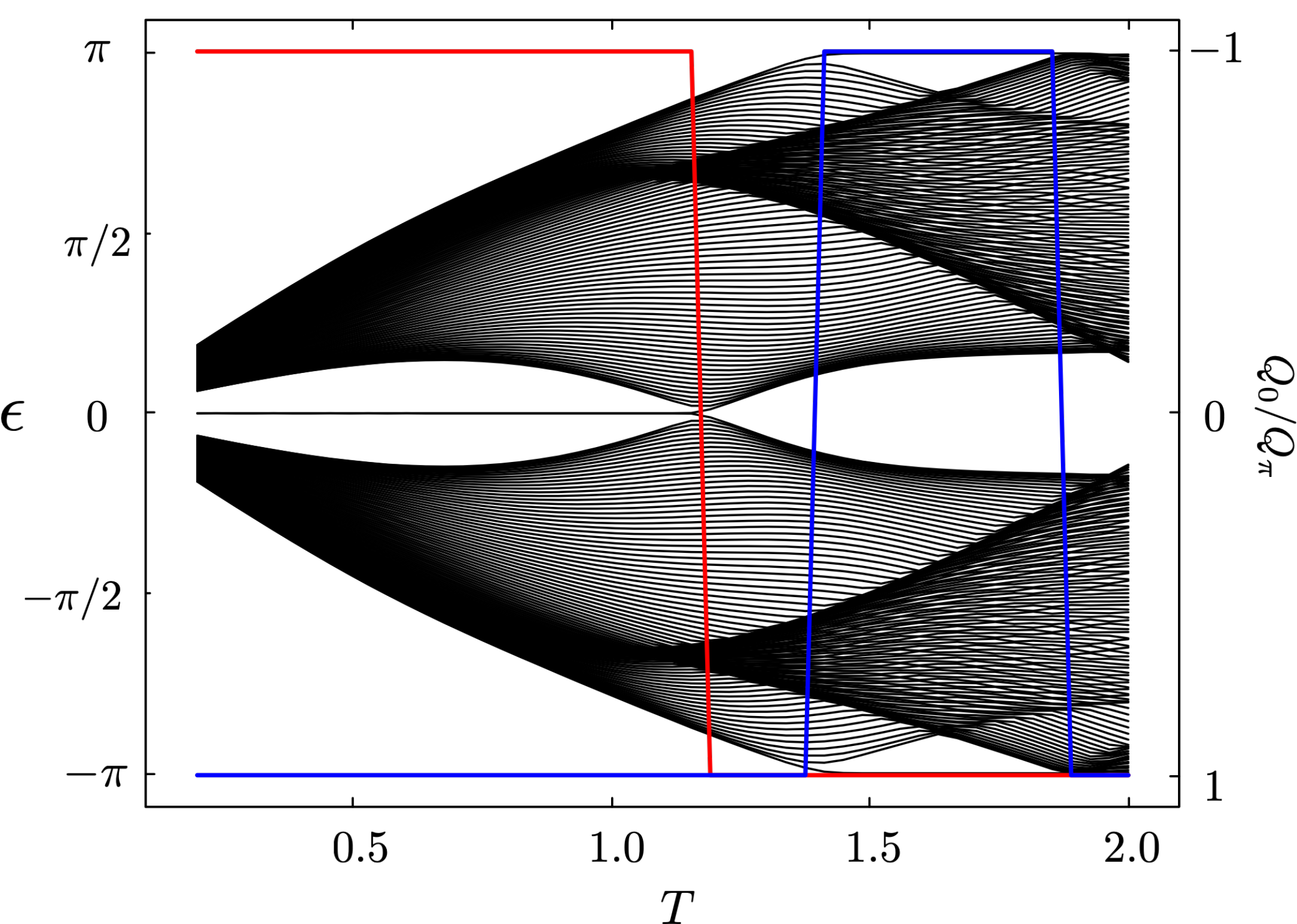}
\caption{\label{spectrum_mod_kitaev}(color online). (black curves) quasienergy spectrum $\epsilon$ for various driving periods of the modulated Kitaev model; (red curves) topological charge for MZM, $Q_{0}$; (blue curves) topological charge for $\pi$ Majorana modes, $Q_{\pi}$.}
\end{figure}

\section{Majorana \texorpdfstring{$\pi$}{pdfbookmark} mode}
In the presence of periodic drive, the energy spectrum becomes periodic in the energy direction. Similar to the Brillouin zone in the space periodic lattice, where the spectrum is periodic in crystal momentum, here we can also define the Floquet Brillouin zone. At the Floquet Brillouin zone boundary, the bands hybridize and can open a topological gap \cite{jiang2011majorana}. In the current context, it can stabilize a new type of Majorana mode, called Majorana $\pi$ mode (MPM) \cite{jiang2011majorana, Liu2013floquet, thakurathi2013floquet, ho2014topological, wang2017line, liu2019floquet}. The condition for Majorana modes is $\gamma^\dagger (E)=\gamma (-E+m\omega)$, which means that these modes can only appear at $E=0$ or $E=\pm \omega/2$. Importantly, the Majorana modes at $E=\pm \omega/2$ are unique to the driven system. In the previous section, we focused on the small $\omega$ region, where the system is mapped to a higher-dimensional system. In this section, we consider a more general $\omega$ region to study the occurrence of MPMs.  We show that our Floquet modulated Kitaev Model and Kondo lattice Model can also host steady unpaired MPMs.

\begin{figure}[b]
\includegraphics[width=85mm]{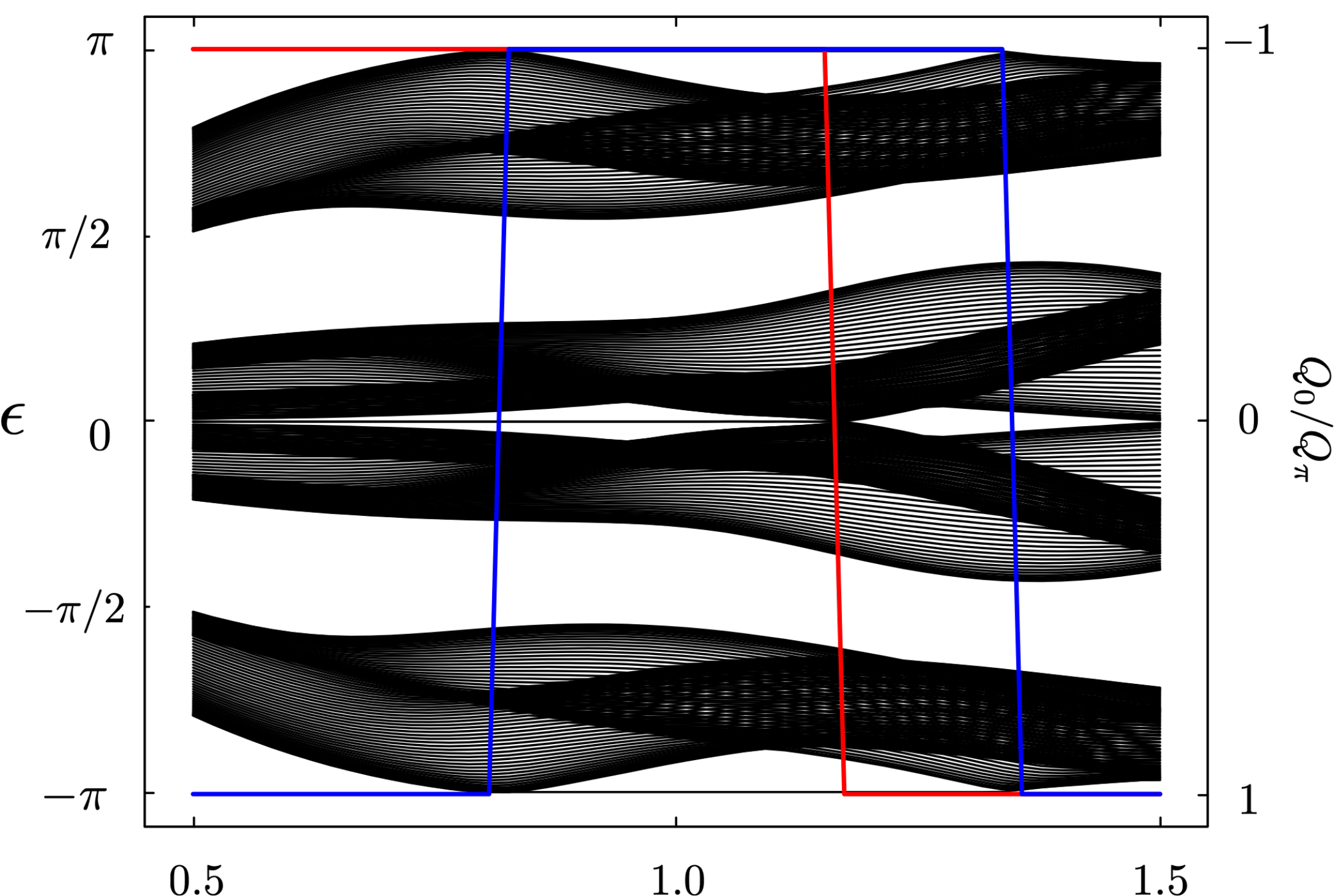}
\caption{\label{spectrum_KLM}(color online). (black curves) quasienergy spectrum $\epsilon$ for various driving periods of Kondo lattice Model; (red curves) topological charge for MZM, $Q_{0}$; (blue curves) topological charge for MPMs, $Q_{\pi}$.}
\end{figure}

We start with the modulated Kitaev model, whose Hamiltonian is in Eq.~\eqref{kitaev}. To calculate the quasienergy, we evaluate the time evolution of the Hamiltonian with open boundary condition after one driving period, $T$,  using $U_{T} = e^{-i\int H_{\rm{K}} dt} \equiv e^{-iH_{\rm{eff}}T}$, and then diagonalize the effective Floquet Hamiltonian $H_{\rm{eff}}$. In this case, we choose $t=\Delta=\mu_0=1.0$, $\delta\mu=4.0$, and $q=2\pi/3$ to calculate the quasienergy spectrum for $T\in \left[ 0.2, 2.0 \right]$. The quasienergy spectrum shows that the modulated Kitaev model hosts MZMs at quasienergy $\epsilon=0$ and MPMs at $\epsilon=\pm\pi/T$ for some different periods, respectively. The topological charges, $Q_{0}=-1$ and $Q_{\pi}=-1$ for the occurrence of MZMs and MPMs respectively, are calculated using the following closed forms \cite{jiang2011majorana}:
\begin{equation}
	Q_{0}Q_{\pi} = \rm{Pf}\left[ M_{0} \right]\rm{Pf}\left[ M_{\pi} \right]\quad Q_{0} = \rm{Pf}\left[ N_{0} \right]\rm{Pf}\left[ N_{\pi} \right]
\label{topo_charge}
\end{equation}
where $M_{p} = \log(U_{T,p})$ and $N_{p} = \log(\sqrt{U_{T,p}})$ are skew-symmetric matrices associated with the time evolution operator at momentum $p$, and $\rm{Pf}\left[M\right]$ is the Pfaffian of matrix $M$. [For more details of the calculations of these topological charges, see Appendix~\ref{app:pimodes}.] The results are summarized in Fig.~\ref{spectrum_mod_kitaev}. In the region $0<T<1.2$, there exists a pair of conventional $\epsilon=0$ MZMs, which is consistent with $Q_0=-1$. In this high frequency region, the system is described by an effective Hamiltonian, which can be obtained from the Magnus high frequency expansion \cite{casas2001floquet, mananga2011introduction}. The $\epsilon=0$ MZMs disappear at $T\approx 1.2$, where the two bands hybridize at $\epsilon=0$ and then open a topologically trivial gap. Around $T\approx 1.45$, the bands hybridize at the Floquent zone boundary, and open a topological gap there. As a consequence, MPMs appear associated with $Q_\pi=-1$.  \par

We also consider the capability for hosting the MPMs in the Kondo lattice Model by adding a Zeeman field to it, whose Hamiltonian now reads:

\begin{equation}
\setlength\abovedisplayskip{0pt}
\begin{aligned}
   \mathcal{H}_{\rm{KL}}=&\sum_{j,\sigma}\left[-t\,c^{\dag}_{j\sigma}c_{j+1\sigma}+|\Delta|c^{\dag}_{j\uparrow}c^{\dag}_{j\downarrow}
           -\mu c^{\dag}_{j,\sigma}c_{j,\sigma}+h.c.\right]\\
           &-\sum_{j,\alpha,\beta}Jc^{\dag}_{j\alpha}(\vec{S_{j}} \cdot \vec{\sigma})_{\alpha\beta}c_{j\beta}-\sum_{j,\alpha, \beta}Bc_{j\alpha}^{\dag}\hat{\sigma}^{z}_{\alpha \beta}c_{j\beta}
\label{KLM_periodic}
\end{aligned}
\end{equation}
where we have promoted $\varphi \rightarrow \omega t$ in $\vec{S_{j}}$ to make the Hamiltonian to be time-dependent. By choosing a set of parameters: $t=\Delta=1.0$, $\mu=0.8$, $J=4.0$, $b=0.5$, $B=2.5$, and $q=2\pi/3$, we repeat the same process mentioned above to find the quasienergy spectrum and the corresponding topological charges for both MZMs and MPMs in periods $T\in \left[ 0.5, 1.5 \right]$. The results are summarized in Fig.~\ref{spectrum_KLM}. Similarly, MZMs and MPMs appear and disappear during the gap opening and closing transitions at $\epsilon=0$ and the Floquet zone boundary.

\section{Conclusion and outlook}

In summary, we demonstrate that the $p_x+ip_y$ superconductivity can be realized in the synthetic dimension by periodically driving the Kitaev model. We also show that a new kind of Majorana $\pi$ mode can appear at the boundary of the Floquet zone in different regions of drive frequency. In both cases, we use the Kondo lattice model, which is a candidate for realizing the Kiteav p-wave chain, to explicitly show the emergence of $p_x+ip_y$ superconductivity and Majorana $\pi$ modes. The Kondo lattice model may be realized in certain heavy fermion compounds \cite{Thompson_Fisk_2012}, and rare earth carbided superconductor \cite{Bulaevskii_1985, Budko_Canfield_2006}. The sliding motion of the magnetic spiral can be achieved by biasing the system with a thermal gradient. Therefore, our results illuminate that the periodic driving of the Kitaev model and the Kondo lattice model provides a fruitful route to realizing new topological phases and excitations that are hard or even impossible to realize in static systems. \par

Our approach to realize synthetic $p_x+ip_y$ superconductivity requires an emergent one dimensional lattice of MZMs, which can be realized in Kondo lattice model with an ellitpical magnetic spiral. Hybridization between the MZMs in the emergent lattice can be tuned by varying their separation through tuning the size of the magnetic domains. For example, consider the case when the single unit cell contains two MZMs, the low energy effective Hamiltonian that describes these MZMs can be written in term of the MZM operators
\begin{equation}
\begin{aligned}
	\mathcal{H}_{t_1-t_2}=\sum_{j}&\left(i\,t_1\gamma_{2j}\gamma_{2j+1}+i\,t_2\gamma_{2j+1}\gamma_{2j+2}\right).
\end{aligned}
\end{equation}
The MZM lattice realized in Kondo lattice model therefore is a platform for studying the interaction effect of MZMs \cite{PhysRevB.92.235123,PhysRevB.91.165402,Rahmani_Franz_2019}, and also the supersymmetry of MZMs proposed in Ref.~\cite{Vish13}.

\begin{acknowledgements}
The work at LANL (SZL) was carried out under the auspices of the U.S. DOE NNSA under contract No. 89233218CNA000001 through the LDRD Program, and was supported by the Center for Nonlinear Studies at LANL (LYY), and was performed, in part, at the Center for Integrated Nanotechnologies, an Office of Science User Facility operated for the U.S. DOE Office of Science, under user proposals $\#2018BU0010$ and $\#2018BU0083$. The work at Virginia was supported by the Center for Materials Theory as a part of the Computational Materials Science (CMS) program, funded by the U.S. Department of Energy, Office of Science, Basic Energy Sciences, Materials Sciences and Engineering Division. The authors also acknowledge the support of Research Computing at the University of Virginia.

\end{acknowledgements}

\appendix
\section{\label{app:rotation}Rotations of Local Basis}
For a general magnetic spiral,
\begin{equation*}
\vec{S}_{j}=(\sin \theta_{j}\cos \phi_{j},\ \sin \theta_{j}\sin \phi_{j},\ \cos \theta_{j})
\end{equation*}
one can define a local rotation in the following form:
\begin{eqnarray*}
   \left(
   \setlength{\arraycolsep}{1.2pt}
 \begin{array}{c}
   c_{j, \uparrow}   \\[6pt]
   c_{j, \downarrow} \\
   \end{array}
   \right)
   =U_{j}
 \left(
   \setlength{\arraycolsep}{1.2pt}
 \begin{array}{c}
   \overline{c}_{j, \uparrow}   \\[6pt]
   \overline{c}_{j, \downarrow} \\
   \end{array}
   \right)
\end{eqnarray*}
where
\begin{eqnarray*}
U_{j}=
   \left(
   \setlength{\arraycolsep}{1.2pt}
 \begin{array}{cc}
   \cos \frac{\theta_{j}}{2} & -\sin \frac{\theta_{j}}{2} e^{-i\phi_{j}}\\[6pt]
   \sin \frac{\theta_{j}}{2} e^{i\phi_{j}} & \cos \frac{\theta_{j}}{2} \\
 \end{array}\right)
\end{eqnarray*}
Then the Hamiltonian in Eq. \eqref{KLM} can be written as:
\begin{equation}
\begin{aligned}
   \mathcal{H}=&\sum_{j,\alpha, \beta}\left(-t\,\overline{c}^{\dag}_{j\alpha}\Omega_{j, \alpha, \beta}\overline{c}_{j+1\beta}+h.c. -J\,\overline{c}^{\dag}_{j\alpha}\sigma^{z}_{\alpha \beta}\overline{c}_{j\beta}\right)\\[5pt]
   -&\sum_{j, \sigma} \mu\, \overline{c}^{\dag}_{j\sigma}\overline{c}_{j\sigma}
   +\sum_{j}\left( \Delta\, \overline{c}^{\dag}_{j\uparrow}\overline{c}^{\dag}_{j\downarrow}
   +h.c.\right)
\end{aligned}
\end{equation}
in which
\begin{eqnarray*}
\Omega_{j}=U^{\dag}_{j}U_{j+1}=
   \left(
   \small {
   \setlength{\arraycolsep}{1.6pt}
 \begin{array}{cc}
   \alpha_{j} & -\beta^{*}_{j} \\[10pt]
   \beta_{j} & \alpha^{*}_{j} \\
 \end{array}
 }\right)
\end{eqnarray*}
and the components are
\begin{equation*}
   \alpha_{j}=\cos\frac{\theta_{j}}{2} \cos\frac{\theta_{j+1}}{2} + \sin\frac{\theta_{j}}{2} \sin\frac{\theta_{j+1}}{2} e^{-i(\phi_{j} - \phi_{j+1})}
\end{equation*}
\begin{equation*}
   \beta_{j}=-\sin\frac{\theta_{j}}{2}  \cos\frac{\theta_{j+1}}{2} e^{i\phi_{j}}+\cos\frac{\theta_{j}}{2} \sin\frac{\theta_{j+1}}{2} e^{i\phi_{j+1}}
\end{equation*}
For our Kondo lattice model with elliptical magnetic spiral, we have $\vec{S_{j}}(\varphi)=(0, -\sin(qj+\varphi), b\cos(qj+\varphi))$ with $b\in (0,1)$. The rotation matrix is reduced to
\begin{eqnarray*}
U_{j}=
   \left(
 \begin{array}{cc}
   \cos\frac{\theta_{j}}{2}   & -i\sin\frac{\theta_{j}}{2} \\[10pt]
   -i\sin\frac{\theta_{j}}{2} & \cos\frac{\theta_{j}}{2}   \\
 \end{array}
 \right)
\end{eqnarray*}
with $\theta_{j}$
\begin{equation}
   \tan\theta_{j}=\frac{\sin(qj+\varphi)}{b\cos(qj+\varphi)}
\end{equation}

This rotation of local basis transforms the local exchange coupling term into:
\begin{eqnarray}
\begin{aligned}
&J\,U^{\dag}_{j}\, \vec{S_{j}} \cdot \vec{\sigma} \,U_{j} \\
=\, &J\,\sqrt{b^2\cos^2(qj+\varphi)+\sin^2(qj+\varphi)}\,\sigma^{z}
\end{aligned}
\end{eqnarray}

\section{\label{app:wstransform}Schrieffer-Wolff Transformation}
When $\widetilde{J}(b,j,\varphi)\approx|\mu|\gg t,\Delta$, and $\mu>0$, we see that near the Fermi level, only the down-spin bands are important. Then, we can perform the Schrieffer-Wolff transformation to project out the up-spin bands, and we obtain the effective Kitaev model \cite{choy2011majorana}. The details of this projection are listed below.

From Eq. \eqref{KLM_periodic}, we can perform a canonical transformation on the Hamiltonian $\widetilde{\mathcal{H}}_{\rm{eff}}=e^{-iS}\widetilde{\mathcal{H}}e^{iS}$ by choosing
\begin{equation}
\begin{aligned}
   S
   =-\sum_{i}\left[ \frac{t\beta_{i}^{*}}{2J}\left( \overline{c}^{\dag}_{i+1\uparrow}\overline{c}_{i\downarrow}-\overline{c}^{\dag}_{i\uparrow}\overline{c}_{i+1\downarrow} \right)\right.
   \left.+ \frac{\Delta}{2\mu}\overline{c}_{i\downarrow}\overline{c}_{i\uparrow} - h.c.\right]
\end{aligned}
\end{equation}
This eliminates the first order matrix elements between the spin-up and spin-down bands. The resulting effective Hamiltonian, $\widetilde{\mathcal{H}}_{\rm{eff}}=\mathcal{H}_{\rm KL}+[\mathcal{H}_{\rm KL}, iS]$, to the first order in $t$ and $\Delta$, is written as:
\begin{equation}
\begin{aligned}
   \widetilde{\mathcal{H}}_{\rm{eff}}=-\sum_{j}&\left[\left(\mu-\widetilde{J}(b,j,\phi)\right)\psi^{\dag}_{j}\psi_{j} +t\alpha_{j}\psi^{\dag}_{j}\psi_{j+1}\right. \\ &+\left.t\Delta\beta^{*}_{j}\left(\frac{1}{2\mu}+\frac{1}{2J}\right) \psi_{j}\psi_{j+1} + h.c.\right]
\end{aligned}
\end{equation}
where we have replaced the \{$\overline{c}_{j\downarrow}$\} operators by \{$\psi_{j}$\}. Thus, the Kondo lattice model is reduced to the Kitaev model in the low-energy limit.

\section{\label{app:pimodes}\texorpdfstring{$Z_{2}\times Z_{2}$}{pdfbookmark} Topological Charges}
The topological charges of the zero and $\pi$ Majorana modes, first introduced in Ref. \cite{jiang2011majorana}, are listed in Eq. \eqref{topo_charge}. The physical meaning of $Q_{0}$ ($Q_{\pi}$) is the parity of the total number of times the corresponding quasienergies of $U_{T,0}$ and $U_{T,\pi}$ cross $\epsilon=0$ ($\epsilon=\pi$) \cite{jiang2011majorana}. In other words, $Q_{0}$ ($Q_{\pi}$) counts the parity of the total number of times that both quasienergies $\epsilon(p=0)$ and $\epsilon(p=\pi)$ cross the zero quasienergy level ($\pi$ quasienergy level). In this section, we briefly discuss the details of the calculation of the topological charges in the modulated Kitaev model. To begin with, one needs to calculate the time evolution matrix $U_{T,p}$ for $p=0$ and $p=\pi$. Then, after directly calculating $M_{p}=\log(U_{T,p})$, we need to perform the transformation such that $M_{p}$ becomes skew symmetric. After that, using $Q_{0}Q_{\pi} = \rm{Pf}\left[ M_{0} \right]\rm{Pf}\left[ M_{\pi} \right]$, we find the result of $Q_{0}Q_{\pi}$.\par

To calculate $Q_{0}$ at a specific $T$, however, in addition to the calculations of the time evolution matrix mentioned above, one needs to diagonalize the effective Hamiltonian, $\overline{H}_{\rm{eff}}=i\log(U_{T,p})$. Performing $P^{-1}\overline{H}_{\rm{eff}}P=D_{T,p}$, where $P$ is the matrix composed of eigenvectors of $\overline{H}_{\rm{eff}}$, we find the diagonal matrix, $D_{T,p}$, composed of the eigenvalues of $\overline{H}_{\rm{eff}}$. It is important to note that, for $t\in[0, T]$, if quasienergy $\epsilon_{T,p}$ crosses $\pi$ odd number times, then there is a jump in quasienergy due to the particle-hole symmetry of the system. Thus, one needs to keep track of the history of time evolution within $t\in[0, T]$ and count the number of times that the quasienergy crosses $\pi$ level. If it is an odd number, one needs to add (subtract) $2\pi$ to the lowest (highest)  quasienergy. After that, one simply performs $P(i\,D_{T,p}/2)P^{-1}$ and a transformation to make it skew-symmetric to find $N_{p}=\log(\sqrt{U_{T,p}})$. Then, using $Q_{0} = \rm{Pf}\left[ N_{0} \right]\rm{Pf}\left[ N_{\pi} \right]$, one can calculate the topological charge for the zero modes. Finally, once we find the result of $Q_{0}$, we can calculate $Q_{\pi}$ via $Q_{0}Q_{\pi}$ mentioned earlier.\par

\bibliography{ref}
\end{document}